\documentclass[usenatbib]{mn2e}
\usepackage{graphicx}
\usepackage{natbib}
\usepackage{amssymb}
\usepackage{aas_macros}

\title[Multi-component dwarf galaxies]{Multiple dynamical components in Local Group dwarf spheroidals}

\author[McConnachie, Pe{\~n}arrubia \& Navarro] {Alan W. McConnachie, Jorge Pe{\~n}arrubia \& Julio F. Navarro\\
Department of Physics and Astronomy, University of Victoria, Victoria, B.C., V8P 1A1, Canada}

\begin{document}

\maketitle

\begin{abstract}
The dwarf spheroidal (dSph) satellites of the Local Group have long
been thought to be simple spheroids of stars embedded within extended
dark matter halos. Recently, however, evidence for the presence of
spatially and kinematically distinct stellar populations has been
accumulating.  Here, we examine the influence of such components on
dynamical models of dwarf galaxies embedded in cold dark matter
halos. We begin by constructing a model of Andromeda~II, a dSph
satellite of M31 which shows evidence for spatially distinct stellar
components. We find that the two-component model predicts an overall
velocity dispersion profile that remains approximately constant at
$\sim 10 - 11$\,km\,s$^{-1}$ out to $\sim 1$\,kpc from the center;
this is despite wide kinematic and spatial differences between the two
individual components. This prediction can be validated by detailed
spectroscopic analysis of this galaxy.  The presence of two components
may also help to explain oddities in the velocity dispersion profiles
of other dSphs; we show that velocity dispersion profiles which appear to rise
from the center outwards before leveling off---such as those of Leo I,
Draco, and Fornax---can result from the gradual transition from a
dynamically cold, concentrated component to a second, hotter, and more
spatially extended one, both in equilibrium within the same dark
halo. Dwarf galaxies with two stellar components generally have a
leptokurtic line-of-sight velocity distribution which is well
described by a double Maxwellian. This may be contrasted with other
dynamical explanations such as a radially-dependent anisotropy in the
stars' orbits. Interestingly, we find that multiple equilibrium
components could also provide a potential alternative origin for
``extra-tidal'' stars (normally ascribed to tidal effects) in
situations where corroborating evidence for tides --- such as
elongation of the main body of the dwarf in the orbital direction or
velocity gradients across its face driven by protruding tidal tails
--- may be lacking.
\end{abstract}

\begin{keywords}
galaxies: dwarf --- galaxies: halos --- galaxies: kinematics and dynamics --- Local Group --- galaxies: structure
\end{keywords}

\section{Introduction}

Dwarf spheroidal (dSph) galaxies are extremely faint systems whose
relatively large size and velocity dispersion combine to make them
premier examples of systems likely dominated by dark matter. They
contain little or no gas, and individual stars are the only available
tracer of their structure and kinematics. Although long
regarded as relatively simple, single-component spheroids of stars,
this view has been challenged by the advent of panoramic imaging
techniques and multi-object spectrographs in large telescopes.

The analysis of the color-magnitude diagrams (CMDs) of dSphs shows, in
many systems, compelling evidence for multiple and protracted episodes
of star formation (eg. \citealt{grebel1997,mateo1998a}, and references
therein), as well as for the presence of spatially distinct stellar
populations (\citealt{dacosta1996,harbeck2001}). These populations
often spill beyond the nominal ``tidal radius'' estimated from
King-model fits to the inner surface brightness profile
(eg. \citealt{munoz2006} and references therein), a result which has
brought into focus questions regarding the role of Galactic tides in
limiting the spatial extent of dSphs, as well as the mass and spatial
extent of their surrounding dark matter halos
(\citealt{johnston2002,penarrubia2007a}).

Spectroscopic studies have also uncovered remarkable complexity in the
dynamics of dSphs. As sufficient data become available, a complex
picture has arisen where kinematic oddities such as cold clumps
(\citealt{kleyna2003}), chemodynamically distinct stellar populations
(\citealt{tolstoy2004,battaglia2006,ibata2006}), and hints of rising
dispersion profiles
(\citealt{wilkinson2004,munoz2005,koch2006a,walker2006a}), are common.
This is at odds with the ``natural'' expectation for well-mixed
stellar systems, where the velocity distribution is expected to be a
smooth function of radius.  Observations thus suggest that it may be
time to relax the standard assumption that dSphs are well described by
a single well-mixed stellar system embedded within a dominant dark
matter halo.

In \cite{penarrubia2007a}, we examined the observational properties of
single component stellar systems (described by King models) embedded
in massive cold dark matter CDM) haloes (described by Navarro, Frenk
\& White (1996, 1997, hereafter NFW) profiles), and applied these
models to the known population of Local Group dSphs. We examine here
the general effect of allowing for multiple stellar components in
these dynamical models. We begin in \S 2 by motivating the modeling
technique using the case of Andromeda~II (And~II; a dSph companion of
M31), and then examine in general the velocity dispersion and surface
density profiles of two-component stellar systems embedded within a
CDM halo. We conclude in \S3 with a discussion of our results and
their wider implications.

\begin{figure*}
  \begin{center}
    \includegraphics[angle=270, width=11.5cm]{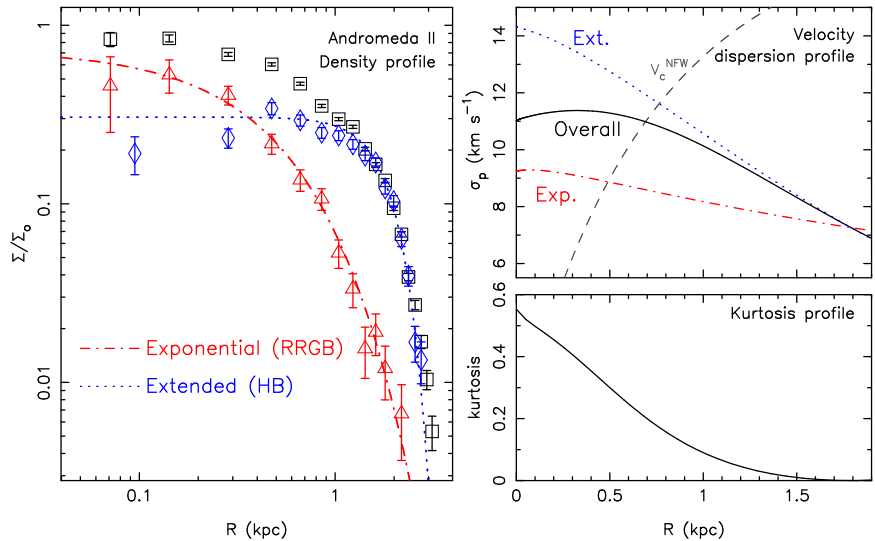}
    \caption{Left panel: the density profile of Andromeda~II derived
    by McConnachie, Arimoto \& Irwin (2007) (black squares). This is
    consistent with the sum of two spatially distinct stellar
    components; an extended component traced by horizontal branch
    stars (HB, blue diamonds/ dotted line) and an exponential
    component traced by the reddest red-giant branch stars (RRGB, red
    triangles/dot-dashed line).  Upper right panel: The predicted
    projected velocity dispersion profile for Andromeda~II (solid
    curve), together with the velocity dispersion profiles of the
    individual components, shown by dot-dashed and dotted
    curves. Lower right panel: the kurtosis of the line-of-sight
    velocity distribution as a function of radius for this
    model. Dwarfs with multiple equilibrium stellar components are
    expected to have a leptokurtic velocity distribution at all radii
    where both components contribute significantly.}
  \end{center} \label{fig:AndII}
\end{figure*}

\section{Dynamics of multiple stellar components}

\subsection{The case of Andromeda II}

And~II is a dSph satellite of M31 located at a distance of $\sim
650$\,kpc from the Milky Way
(\citealt{mcconnachie2004a,mcconnachie2005a}).
\cite{mcconnachie2006b} present its surface brightness profile based
on Isaac Newton Telescope wide field imaging, and suggest that a
discontinuity in the profile at $r \sim 2^\prime$ may be evidence that
it possesses a secondary ``core'' component.

Deeper, follow-up imaging of And~II was obtained by
\cite{mcconnachie2007a} using the Subaru Suprime-Cam wide field
camera. The black squares in Figure~1 reproduce the overall radial
profile of And~II derived using these data. The deeper data allow the
derivation of the radial profiles of distinct stellar populations; the
bottom two profiles shown in the same panel trace the spatial
distribution of horizontal branch stars (HB) and the reddest red-giant
branch (RRGB) stars. Clearly, the HB stars are more spatially extended
than the RRGB stars, which trace an exponential profile. As discussed
in detail in \cite{mcconnachie2007a}, it is very likely that these two
profiles represent the density distribution of two distinct stellar
components in And~II. Not only does an appropriately weighted sum of
these two profiles fits the overall number density profile remarkably
well, as indicated in Figure~1, but it also explains significant
differences seen in the stellar populations at small and large
radii. The reader is referred to \cite{mcconnachie2007a} for more
detail on these data and on the construction of these profiles.

Large differences in spatial distributions of stars should manifest
themselves kinematically, and this can in principle be used to
constrain the properties of the overall potential. As described by
\cite{penarrubia2007a}, one expects a direct link between the
velocity dispersion and the spatial distribution of stars orbiting
within a CDM halo. A stellar component deeply embedded within the
central regions of a CDM halo is expected to have a low central velocity
dispersion and a dispersion profile that remains flat well outside its
characteristic core radius. More extended components, on the other
hand, should have higher central velocity dispersions and more steeply
declining dispersion profiles.

\begin{figure*}
  \begin{center}
    \includegraphics[angle=270, width=11.5cm]{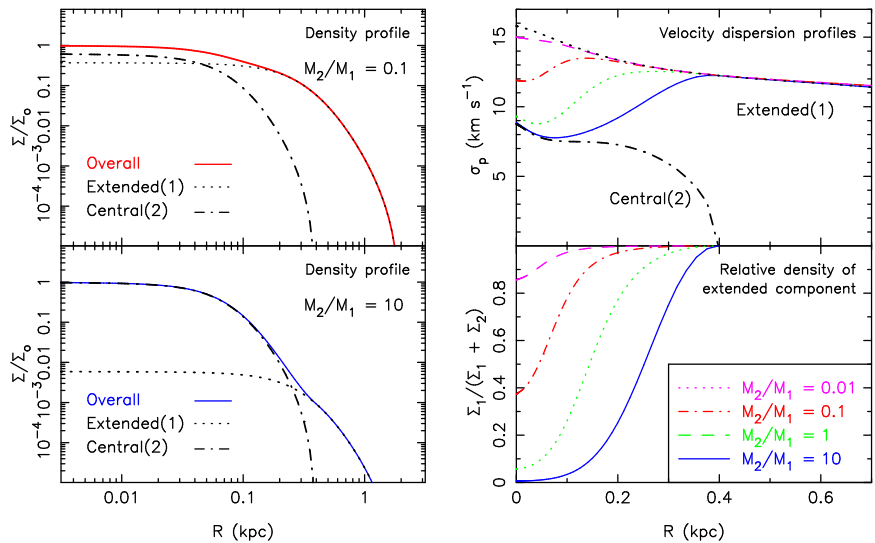}
    \caption{Left panel: Projected density profile of a system
    consisting of the superposition of two King models with mass
    ratios as labeled.  Top right panel: Projected velocity dispersion
    profiles of the two stellar components and the overall projected
    velocity dispersion profile (line styles are as in the bottom
    right panel). Bottom right panel: The relative contribution of the
    more extended component to the total projected stellar density,
    shown as a function of radius for various choices of the mass
    ratio $M_2/M_1$.}
    \label{fig:sigmap}
  \end{center}
\end{figure*}

Detailed kinematic studies of And~II are currently lacking. However,
\cite{cote1999a} measure a central velocity dispersion of $9.3 \pm
3$\,km\,s$^{-1}$ based on 7 stars. Given the expected dominance of the central
component in the inner few hundred parsecs of And~II, it is
likely that this measurement is representative of the central velocity
dispersion of the exponential component found by
\cite{mcconnachie2007a}. Following the technique of
\cite{penarrubia2007a}, we adopt this value and the appropriate value
for the `core' radius of the central component ($r_c \sim 370$\,pc) to
determine the properties of a $\Lambda$CDM halo at $z = 0$ compatible
with these constraints. The resulting NFW profile may be fully
characterized by the peak of the circular velocity curve, $(V_{\rm
max},r_{\rm max})$: we find $(V_{\rm max} = 17.4$\,km\,s$^{-1}, r_{\rm
max} = 3.06$\,kpc$)$.

The velocity dispersion profiles of both components can then be
computed, assuming isotropy, by solving Jeans' equations separately
for each component within the NFW potential. The predicted velocity
dispersion profiles for And~II are shown in the top right panel of
Figure~1. As expected, the exponential component has a lower central
velocity dispersion than the extended component and its profile
declines by only $\sim 1$\,km\,s$^{-1}$ out to $R \sim 1$ kpc. The
more extended component, on the other hand, has a steeply declining
profile, dropping from a central value of $\sim 14.3$\,km\,s$^{-1}$ to
$\sim 7$\,km\,s$^{-1}$ at $R \sim 2$ kpc. For comparison, the circular
velocity profile of the NFW halo is plotted as a gray dashed line as a
function of actual, not projected, radius.

The overall velocity dispersion profile of And II is given by the
weighted sum of the profiles of the components. It has a central
velocity dispersion of $\sim 11$\,km\,s$^{-1}$ which rises slightly
and then declines slowly; at $R \sim 1$\,kpc, the velocity dispersion
is $\sim 10$\,km\,s$^{-1}$, and at $R \sim 2$\,kpc the dispersion has
dropped to $\sim 7$\,km\,s$^{-1}$. Varying the adopted values of
$(V_{\rm max},r_{\rm max})$ by using the different constraints
discussed in \cite{penarrubia2007a} does not alter the central
velocity dispersions significantly. The general shape of the overall
profile of And~II is similar to that expected for a one component
galaxy, but here it is a result of the varying radial contribution of
two dynamical components; this is very distinct from the standard
interpretation of a flat dispersion profile arising from only one
stellar component

Our predictions for And~II are verifiable using current generations of
multi-object spectrographs by obtaining spectroscopic metallicities of
red giant branch stars in And~II.  \cite{mcconnachie2007a} conclude
that the two components of And~II have distinct metallicities, and so
it should be possible to divide the red giant branch stars based on
their metallicity. This technique has been successfully
employed by \cite{tolstoy2004} and \cite{battaglia2006} to uncover
kinematically distinct components in Sculptor and Fornax,
respectively. Therefore, the model which we propose for And~II is
eminently falsifiable.

One may also resort to the {\it distribution} of line-of-sight
velocities. If And II is largely made up of two independent
equilibrium components of widely different dispersion, the velocity
distribution will approach a double Maxwellian distribution whose
kurtosis ($\mu_4/\mu_2^2 - 3$) will be positive (leptokurtic)
everywhere where both components contribute in comparable amounts to
the density profile. This is shown in the bottom right panel of
Figure~\ref{fig:AndII}, where the kurtosis peaks at the center and
declines gradually in the outer regions as the contribution from the
exponential component drops. The uncertainty in the kurtosis
measurement is $\sqrt{24/N}$, where $N$ is the number of stars in the
sample. Thus to measure the predicted value of the kurtosis in the
central regions of And~II to $\sim 3-\sigma$ will require of order 200
stars; large kinematic sample are crucial for accurate kurtosis
measurements in dwarf galaxies.

\subsection{Two-component systems}

As a more general illustration of the interplay between $\sigma_p(R)$
and multiple components, we show in the left panels of Figure~2 the
density profile corresponding to two King models embedded within an
NFW halo. We use King models to describe the stellar distributions
here because they are a convenient means of parameterising the density
profile and have been used extensively is the literature; no
additional physics is implied by their usage.  We set the core and
tidal radii of the more extended component to $r_{c_1} = 400$\,pc,
$r_{t_1} = 2000$\,pc (dotted lines), and $r_{c_2} = 100$\,pc, $r_{t_2}
= 400$\,pc (dot-dashed lines) for the more compact one. The mass ratio
of the two components varies from $M_2/M_1 = 0.1$ in the top panel to
$M_2/M_1 = 10$ in the bottom panel. The overall stellar density
distribution is shown as a solid line in each panel; that of the top
panel is unlikely to be observationally distinguished from a single
component system in the absence of other data; the one in the bottom
panel shows a ``bump'' in the outer density profile, not unlike those
reported in some dSphs (eg. \citealt{irwin1995,sohn2006}).

The top right panel of Figure~2 shows the line-of-sight velocity
dispersion profile, $\sigma_p\left(R\right)$, corresponding to these
two components (embedded in an NFW halo with $r_{\rm max}=2$ kpc and
$V_{\rm max}=30$ km s$^{-1}$).  The bottom right panel shows how the
importance of the hotter, more extended component (\#1) varies with
$R$ for various choices of $M_2/M_1$. Note that, as in the case of And
II, the more extended component (\#1) has a declining velocity
dispersion profile, in contrast with that of the more concentrated one
(\#2), which remains more or less flat all the way to its tidal
radius, $r_{t_2}$.

The detailed shape of the overall velocity dispersion profile depends
on the relative contribution of the two components and on their
velocity difference. The various lines in the top panel of
Figure~\ref{fig:sigmap} illustrate this for various values of the
ratio $M_2/M_1$ (see key in the bottom right panel). When one
component dominates everywhere the velocity dispersion profile is
indistinguishable from a single component model. As variations in the
relative contribution of the two components become more significant,
however, the velocity dispersion profile starts to deviate from that
for a single component system. The exact shape of the resulting
$\sigma_p\left(R\right)$ profile depends closely on
$\Sigma_1/\left(\Sigma_1 + \Sigma_2\right)$; $\sigma_p\left(R\right)$
may therefore set additional constraints on the spatial properties in two
component models.

\begin{figure}
  \begin{center}
    \includegraphics[angle=270, width=8cm]{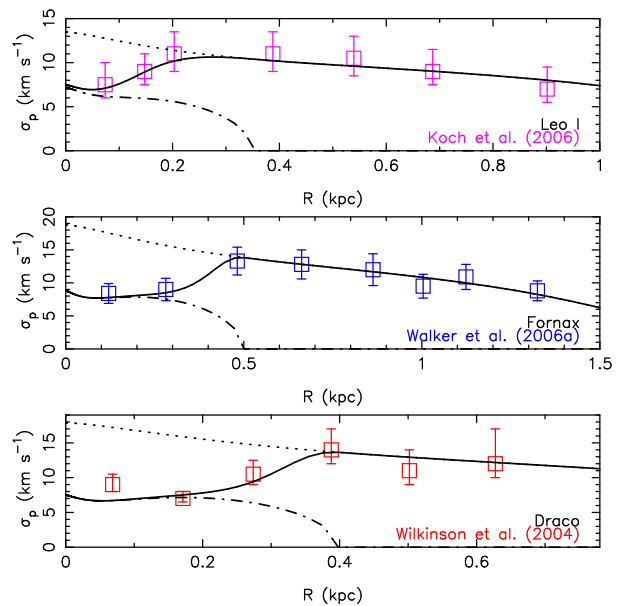}
    \caption{Velocity dispersion profiles of three Local Group dSphs
    which show a rise in their velocity dispersions from the center
    outwards before leveling off, taken from the literature. Possible,
    but degenerate, decompositions of these profiles into a cold and
    hot component are shown to illustrate the general plausibility of
    two dynamical components instead of one.}
    \label{fig:ldf}
  \end{center}
\end{figure}

\section{Discussion}

In two component models, the centrally concentrated component is
dynamically colder and has a decreasing influence on the overall
dispersion profile as a function of radius. In the extreme case where
each component dominates at different radii, we transition from a cold
to a hot velocity dispersion profile. It is intriguing to note that
such features in the velocity dispersion profile may not be unusual in
Local Group dSphs. This is illustrated in Figure~\ref{fig:ldf}, where
we show data for Leo I, Fornax, and Draco, compiled from the
literature. In each of these three cases, $\sigma_p$ appears to rises
by about a factor of $\sim 2$ from its central value before starting
to decline again. In the case of Fornax, we refer the reader to
\cite{battaglia2006}.  The velocity dispersion profile of Draco has
previously been fitted by \cite{mashchenko2006} using a single stellar
component with a radially varying orbital anisotropy embedded in a
massive halo, and \cite{koch2006a} imply a similar anisotropy profile
is required for Leo~I if it resides in an NFW halo.

\cite{klimentowski2006} suggest that that these bumps in the velocity
dispersion profiles are due to tidal effects. However, we consider
this unlikely for two reasons (i) as concluded by \cite{read2006a} and
\cite{klimentowski2006}, tides do not appear to affect dSph satellites
inside of $1$\,kpc, and so are unlikely to produce observable features
at radii of typically $100 - 400$\,pc (ii) the presence of tidally stripped
stars manifest themselves as monotonically increasing features in the
velocity dispersion profiles of dSphs
(\citealt{read2006a,klimentowski2006}). The features highlighted in
Figure~3, however, rise and then level off or decline. Thus we
conclude their origin is unlikely tidal.

For each galaxy in Figure~3 we have shown a possible decomposition of
the velocity dispersion profile into a cold and hot component. These
curves are inevitably degenerate, given that many parameters need to
be fitted to describe two stellar components and a dark matter
halo. In addition, we have not attempted to determine the
uncertainties due to observation errors or bin sizes since these
profiles have been taken direct from the literature.  Some caution is
clearly required; for example, recent data for Leo I do not appear to
show such as pronounced rise in the dispersion profile as found by
Koch et al. (M. Mateo, private communication). As such, these curves
merely demonstrate the {\it plausibility} of multiple components in
these systems and illustrate a few fairly general conclusions that can
be drawn.

The first is that the rise in $\sigma_p$ from the center outwards
reflects the rise in the relative contribution of the more extended
component, a fact that may be used to constrain the relative size of
the two components and to pinpoint the radius where both components
contribute more or less equally to $\sigma_p(R)$.  This is where
deviations from a simple Maxwellian-like velocity distribution are
likely to be maximized. At this position, the line-of sight velocity
distribution would be best characterized by a double Maxwellian
distribution, distinguishing it from other proposals such as a
radially-varying orbital anisotropy (see \cite{mashchenko2006} for
application of this idea to Draco).

The ratio of central-to-maximum velocity dispersion provides
a lower bound to the ratio of the central velocity dispersions of the
two components.  Interpreting Leo I as a two-component King model, the
data shown in Figure~\ref{fig:ldf} suggest that the central
component has a central (projected) velocity dispersion of $\sim
6$--$7$\,km\,s$^{-1}$, and that the outer component has
$\sigma_{p_2}(0) \sim 13$-$14$ km s$^{-1}$.  If Leo I is a multiple
component system, we expect that the line-of-sight velocity
distribution of stars at about $R \simeq 150-200$\,pc should reveal
evidence for the double Maxwellian distribution discussed
previously. Similar testable inferences can be made for Fornax and
Draco.

We note that a multiple component system is not required to exhibit
differences in the stellar populations of its components. For example,
one could build a dwarf by merging two progenitors of very different
spatial extent and kinematics but similar age and metallicity. The
remnant of such a merger would appear homogeneous to a CMD analysis
but would retain the signature of the two progenitors in phase
space. Within this context, evidence for a hierarchical origin of a
dSph might be best appreciated in the dynamical evidence rather than
in the CMDs.

Perhaps the most common deviation from single component models is the
presence of stars that are clearly associated with the dSph, but which
lie beyond the putative limiting radius of a King model fit. The
presence of these ``extra-tidal'' stars can be caused by dynamical
heating in the presence of the tidal field of the Galaxy and has
attracted considerable attention because of its potential to
constrain the mass and extent of a dSph's dark halo (e.g.,
\citealt{johnston2002,read2006a,read2006b,klimentowski2006}).

As illustrated in Figure~2, the presence of a low density, extended
second component, could in some instances be mistaken for
``extra-tidal'' stars, although in this scenario the stars are
actually bound and in equilibrium with the dark matter halo. Note also
that, in this interpretation, we would expect $\sigma_p$ to {\it
decline} in this region, since these stars belong to the extended
component. Thus, a multiple component scenario offers an alternative
explanation for ``extra-tidal'' stars in situations where
corroborating evidence for tides --- such as elongation of the main
body of the dwarf in the orbital direction or velocity gradients
across the face of the dwarf driven by protruding tidal tails --- may
be lacking.

If the multiple component scenario is correct, a natural question
which arises from this discussion is: {\it how did these systems
develop and preserve such complex structures}? A spatially varying
star formation history, where subsequent epochs of star formation
occurred in different volumes than previous ones, has been suggested
by \cite{kawata2006}.  Alternatively, \cite{battaglia2006} suggest
that there is tentative evidence for non-equilibrium kinematics in
their data for Fornax, which might imply a merger-driven
scenario. However, it is currently too soon to distinguish between
these and other scenarios. Given the observational discovery of
multiple structural components in dSph galaxies, it is important that
the consequences for dynamical models of these systems are fully
explored.

\section*{Acknowledgments}

We thank Nobuo Arimoto, Mike Irwin and the DART team for assistance
and useful conversations during this work. AWM is supported by a
Research Fellowship from the Royal Commission for the Exhibition of
1851, and thanks Sara Ellison and Julio Navarro for additional
support. We thanks the referee, Mario Mateo, for useful suggestions
and for sharing his data on Leo I prior to publication.

\bibliographystyle{apj}
\bibliography{/Users/alan/Papers/references}
\end{document}